\documentclass[twoside,leqno,twocolumn]{article}  
\usepackage{ltexpprt} 

\usepackage{subfigure}
\usepackage{epsfig}
\usepackage{amsmath}
\usepackage{amssymb}
\usepackage{graphicx}

\newcommand{\wo}{\setminus}
\newcommand{\setR}{\mathbb{R}}
\newcommand{\plane}{\setR^2}
\newcommand{\fit}{\mathrm{fit}}
\newcommand{\enc}{\mathrm{enc}}
\renewcommand{\epsilon}{\varepsilon}
\newcommand{\calB}{\mathcal{B}}
\newcommand{\calC}{\mathcal{C}}
\newcommand{\calI}{\mathcal{I}}

\newcommand{\calO}{\mathcal{O}}
\newcommand{\IC}{\mathrm{IC}}
\newcommand{\cdist}{\tilde{d}}

\newcommand{\inV}{\in V}

\newcommand{\IMGQUALITY}{lo}

\begin{document}

\title{\Large Deterministic boundary recognition and topology extraction\\
for large sensor networks}
\author{
Alexander Kr{\"o}ller\thanks{Department of Mathematical Optimization, Braunschweig University of Technology, D-38106 Braunschweig, Germany. Email: \{\texttt{a.kroeller}, \texttt{s.fekete}\} \texttt{@tu-bs.de}.}\ \thanks{Supported by
DFG Focus Program 1126, ``Algorithmic Aspects of Large and Complex Networks'',
Grants Fe 407/9-2 and Fi 605/8-2.}
\and S\'andor P.\ Fekete\footnotemark[1]
\and Dennis Pfisterer\thanks{Institute of Telematics, University of L{\"u}beck, D-23538 L{\"u}beck, Germany.\hspace{6cm}\mbox{} Email: \{\texttt{pfisterer}, \texttt{fischer}\} \texttt{@itm.uni-luebeck.de}.}\ \footnotemark[2]
\and Stefan Fischer\footnotemark[3]}
\date{}

\maketitle

\begin{abstract} \small\baselineskip=9pt
  We present a new framework for the crucial challenge of
  self-organization of a large sensor network. The basic scenario can
  be described as follows: Given a large swarm of immobile sensor
  nodes that have been scattered in a polygonal region, such as a
  street network.  Nodes have no knowledge of size or shape of the
  environment or the position of other nodes. Moreover, they have no
  way of measuring coordinates, geometric distances to other nodes, or
  their direction. Their only way of interacting with other nodes is
  to send or to receive messages from any node that is within
  communication range.  The objective is to develop algorithms and
  protocols that allow self-organization of the swarm into large-scale
  structures that reflect the structure of the street network, setting
  the stage for global routing, tracking and guiding algorithms.

  Our algorithms work in two stages: boundary recognition and topology
  extraction. All steps are strictly deterministic, yield fast
  distributed algorithms, and make no assumption on the distribution
  of nodes in the environment, other than sufficient density.
\end{abstract}

\section{Introduction}
\label{sec:intro}

In recent time, the study of wireless sensor networks (WSN) has become
a rapidly developing research area that offers fascinating
perspectives for combining technical progress with new applications of
distributed computing. Typical scenarios involve a large swarm of
small and inexpensive processor nodes, each with limited computing and
communication resources, that are distributed in some geometric
region; communication is performed by wireless radio with limited
range.  As energy consumption is a limiting factor for the lifetime of
a node, communication has to be minimized. Upon start-up, the swarm
forms a decentralized and self-organizing network that surveys the
region.

From an algorithmic point of view, the characteristics of a sensor
network require working under a paradigm that is different from
classical models of computation: absence of a central control unit,
limited capabilities of nodes, and limited communication between nodes
require developing new algorithmic ideas that combine methods of
distributed computing and network protocols with traditional
centralized network algorithms. In other words: How can we use a
limited amount of strictly local information in order to achieve
distributed knowledge of global network properties?

This task is much simpler if the exact location of each node is known.
Computing node coordinates has received a considerable amount of
attention.  Unfortunately, computing exact coordinates requires the
use of special location hardware like GPS, or alternatively, scanning
devices, imposing physical demands on size and structure of sensor
nodes.  As we demonstrated in our paper~\cite{kfb-kl-05}, current
methods for computing coordinates based on anchor points and distance
estimates encounter serious difficulties in the presence of even small
inaccuracies, which are unavoidable in practice.

\begin{figure}
  \centering
  \subfigure[60,000 sensor nodes, uniformly distributed in a polygonal region.\label{fig:city:b}]{
    \includegraphics[height=2.7cm]{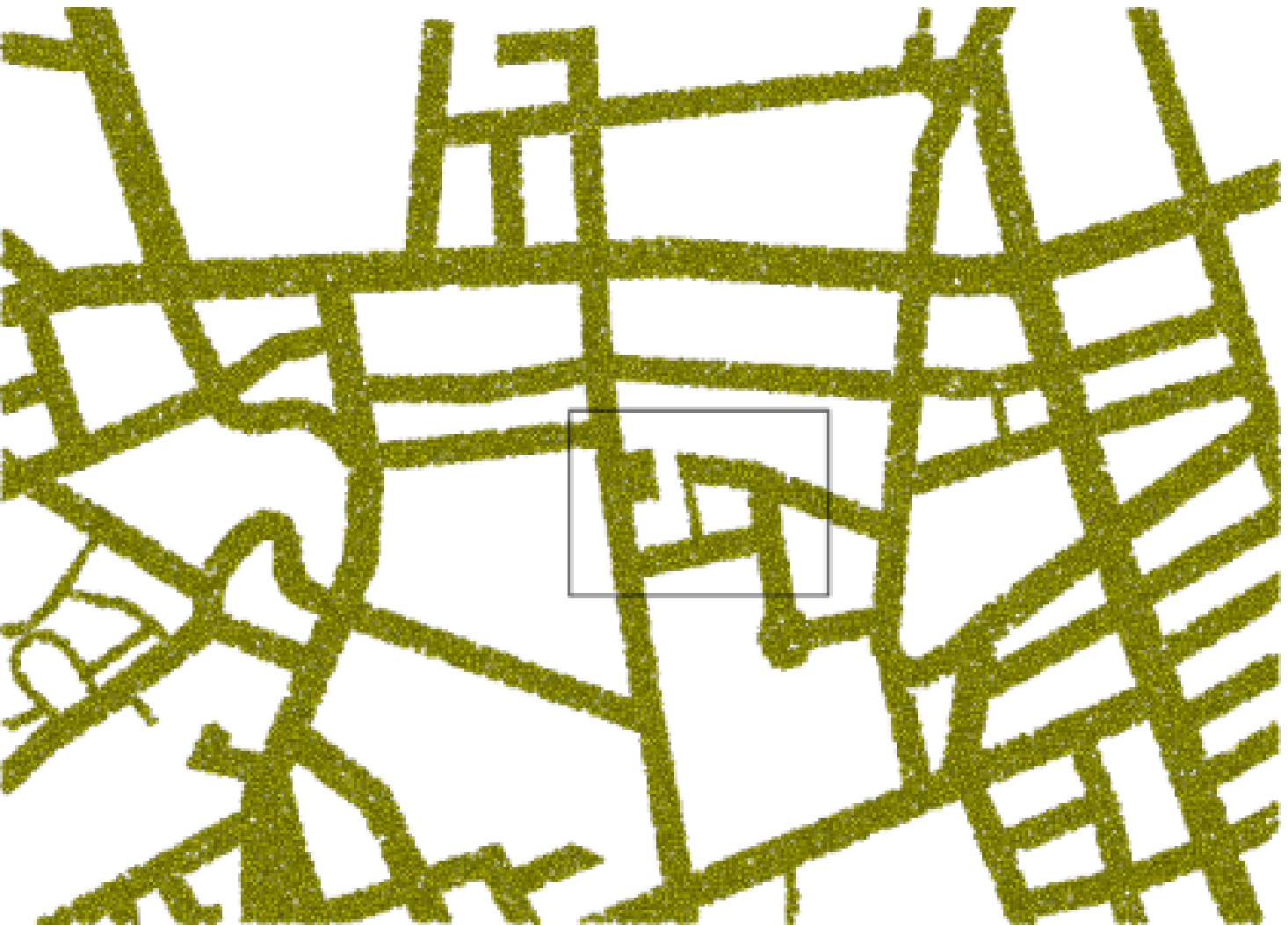}
  }%
  \subfigure[A zoom into (a) shows the communication graph.\label{fig:city:c}]{
    \includegraphics[height=2.7cm]{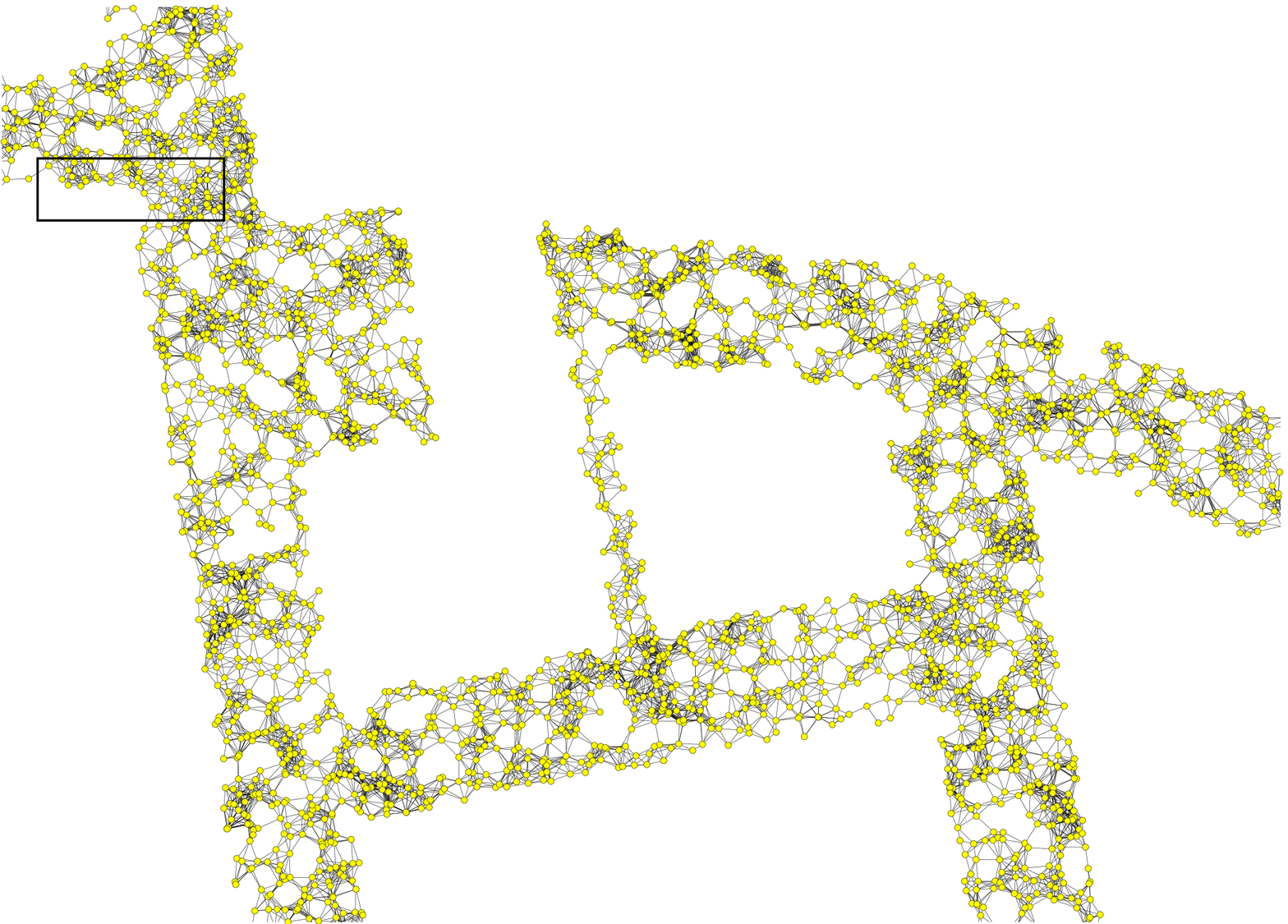}
  }

  \vspace*{-6mm}
  \subfigure[A further zoom shows the communication ranges.\label{fig:city:d}]{
    \makebox[8cm]{\includegraphics[height=2.7cm]{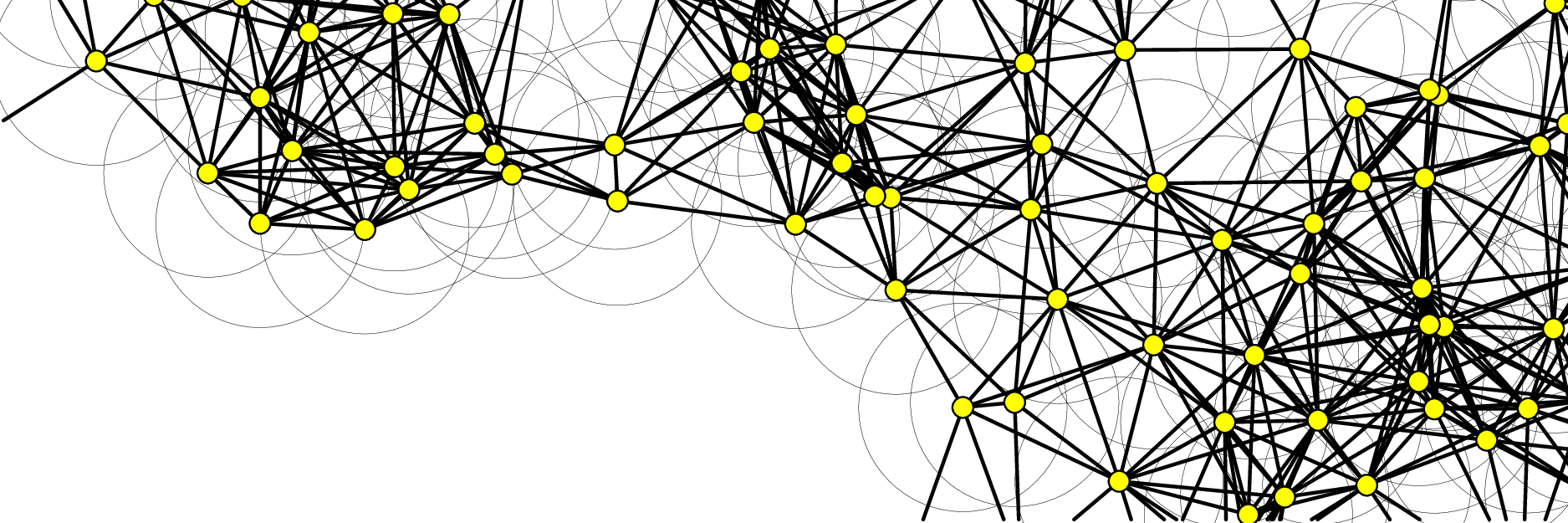}}
  }
  \vspace*{-6mm}
  \caption{Scenario of a geometric sensor network, obtained by scattering sensor nodes in the street network surrounding Braunschweig University of Technology.}
  \label{fig:city}
\end{figure}
When trying to extract robust cluster structures from a huge swarm of nodes
scattered in a street network of limited size, trying to obtain individual
coordinates for all nodes is not only extremely difficult,
but may indeed turn out to be a red-herring chase. 
As shown in \cite{fkp-nbtrsn-04}, there is a way to sidestep many of the above
difficulties, as some structural location aspects do {\em not}
depend on coordinates.
This is particularly relevant for sensor networks
that are deployed in an environment
with interesting geometric features. (See \cite{fkp-nbtrsn-04}
for a more detailed discussion.) Obviously, scenarios as the one
shown in Figure~\ref{fig:city} pose a number of interesting
geometric questions. Conversely, exploiting the basic fact
that the communication graph of a sensor network
has a number of geometric properties provides
an elegant way to extract structural information.

One key aspect of location awareness is {\em boundary recognition},
making sensors close to the boundary of the surveyed region aware of
their position.  This is of major importance for keeping track of
events entering or leaving the region, as well as for communication
with the outside. More generally, any unoccupied part of the region
can be considered a hole, not necessary because of voids in the
geometric region, but also because of insufficient coverage,
fluctuations in density, or node failure due to catastrophic events.
Neglecting the existence of holes in the region may also cause
problems in communication, as routing along shortest paths tends to
put an increased load on nodes along boundaries, exhausting their
energy supply prematurely; thus, a moderately-sized hole (caused by
obstacles, by an event, or by a cluster of failed nodes) may tend to
grow larger and larger. (See \cite{guibas}.)
Therefore, it should be stressed that even though in our basic street
scenario holes in the sensor network are due to holes in the filled
region, our approach works in other settings as well.

Once the boundary of the swarm 
is obtained, it can be used as a stepping stone for extracting
further structures. This is particularly appealing in our scenario, in which
the polygonal region is a street network: In that scenario, we have a 
combination of interesting geometric features, a natural underlying structure
of moderate size, as well as a large supply of practical and relevant
benchmarks that are not just some random polygons, but readily available from
real life.
More specifically, we aim at identifying the graph in which intersections
are represented by vertices, and connecting streets are represented
by edges.  This resulting cluster structure is perfectly suited
for obtaining useful information for purposes like routing, tracking
or guiding. Unlike an arbitrary tree structure that relies
on the performance of individual nodes, it is robust.

\paragraph{Related Work:}
\cite{barriere01qudgrouting} is the first paper to introduce a communication
model based on quasi-unit disk graphs (QUDGs). 
A number of articles deal with node coordinates;
most of the mathematical results are negative, even in a centralized
model of computation.  \cite{breu98unit} shows that unit disk graph (UDG)
recognition is NP-hard, while \cite{aspnescomputational} shows
NP-hardness for the more restricted setting in which all edge lengths 
are known. \cite{kuhn04udgapprox}
shows that QUDG recognition, i.e., UDG approximation, is also hard;
finally, \cite{bruck05anglelocalization} show that UDG embedding is
hard, even when all angles between edges are known.  The first paper
(and to the best of our knowledge, the only one so far) describing an
approximative UDG embedding is \cite{moscibroda04virtualcoordinates};
however, the approach is centralized and probabilistic, yielding (with
high probability) a $\calO(\log^{2.5} n  \sqrt{\log\log n})$-approximation.

There are various papers dealing with heuristic localization algorithms;
e.g., see 
\cite{capkun01gpsfree,doherty01convexpositioning,priyantha03anchorfreelocalization,savarese02robust,sundaram02connectivitylocation}.
In this context, see our paper \cite{kfb-kl-05} for an experimental
study pointing out the serious deficiencies of some of the resulting coordinates.

\paragraph{Main Results:}
Our main result is the construction of an overall framework that
allows a sensor node swarm to self-organize into a well-structured
network suited for performing tasks such as routing, tracking
or other challenges that result from popular visions of what sensor
networks will be able to do.  The value of the overall framework is
based on the following aspects:

\begin{itemize}
\item We give a distributed, deterministic approach for identifying
nodes that are in the interior of the polygonal region, or near its boundary.
Our algorithm is based on topological considerations and geometric packing 
arguments. 
\item Using the boundary structure, we describe a distributed, deterministic
approach for extracting the street graph from the swarm. This module also
uses a combination of topology and geometry.
\item The resulting framework has been implemented and tested
in our simulation environment {\sc Shawn}; we display some experimental
results at the end of the paper.
\end{itemize}

The rest of this paper is organized as follows. In the following Section 2
we describe underlying models and introduce necessary notation. Section 3 
deals with boundary recognition. This forms the basis for topological
clustering, described in Section~4. Section~5 describes some computational
experiments with a realistic network.


\section{Models and Notation}
\label{sec:model}

\paragraph{Sensor network:}
A {\em Sensor Network} is modeled by a graph $G=(V,E)$, with an edge
between any two nodes that can communicate with each other.  For a
node $v\in V$, we define $N_k(v)$ to be the set of all nodes that can
be reached from $v$ within at most $k$ edges. The set $N(v)=N_1(v)$
contains the direct neighbors of $v$, i.e., all nodes $w\in V$ with
$vw\in E$. For convenience we assume that $v\in N(v)$ $\forall v\in
V$. For a set $U\subseteq V$, we define $N_k(U):=\cup_{u\in U}N_k(u)$.
The size of the largest $k$-hop neighborhood is denoted by
$\Delta_k:=\max_{v\in V} |N_k(v)|$. Notice that for geometric radio
networks with even distribution, $\Delta_k=\calO(k^2\Delta_1)$ is a reasonable
assumption.

Each node has a unique ID of size $\mathcal{O}(\log |V|)$. The
identifier of a node $v$ is simply $v$.

Every node has is equipped with local memory of size
$\mathcal{O}(\Delta^2_{\mathcal{O}(1)}\log |V|)$.  Therefore, each node can
store a sub\-graph consisting of nodes that are at most $\calO(1)$ hops
away, but not the complete network.

\paragraph{Computation:}
Storage limitation is one of the main reasons why sensor networks
require different algorithms: Because no node can store the whole
network, simple algorithms that collect the full problem data at some
node to perform centralized computations are infeasible. 

Due to the distributed nature of algorithms, the classic means to
describe runtime complexity are not sufficient. Instead, we use
separate {\em message} and {\em time} complexities: The former
describes the total number of messages that are sent during algorithm
execution. The time complexity describes the total runtime of the
algorithm over the whole network.

Both complexities depend heavily on the computational model. For our 
theoretical analysis, we use a variant of the well-established 
$\mathcal{CONGEST}$ model \cite{peleg00distributedcomputing}: All nodes 
start their local algorithms at the same time ({\em simultaneous wakeup}). 
The nodes are
synchronized, i.e., time runs in {\em rounds} that are the same for
all nodes. In a single round, a node can perform any computation for 
which it has complete data. All messages arrive at the
destination node at the beginning of the subsequent round, even if
they have the same source or destination. There are no congestion or
message loss effects. The size of a message is limited to
$\calO(\log|V|)$ bits. Notice that this does only affect the message
complexity, as there is no congestion. We will use messages of larger
sizes in our algorithms, knowing that they can be broken down into
smaller fragments of feasible size.

\paragraph{Geometry:}
All sensor nodes are located in the two-dimensional plane, according
to some mapping $p:V\to\plane$. It is a common assumption that the
ability to communicate depends on the geometric arrangement of the
nodes. There exists a large number of different models that formalize
this assumption. Here we use the following reasonable model:


We say $p$ is a {\em $d$-Quasi Unit Disk Embedding} of $G$ for
parameter $d\leq 1$, if both
\begin{eqnarray*}
  uv\in E &\Longrightarrow& \|p(u)-p(v)\|_2\leq 1\\
  uv\in E &\Longleftarrow& \|p(u)-p(v)\|_2\leq d
\end{eqnarray*}
hold. $G$ itself is called a {\em $d$-Quasi Unit Disk Graph} ($d$-QUDG)
if an embedding exists.  A $1$-QUDG is called a {\em
  Unit Disk Graph} (UDG). Throughout this paper we assume that $G$ is
a $d$-QUDG for some $d\geq\tfrac{1}{2}\sqrt{2}$. The reason for this
particular bound lies in Lemma~\ref{thm:pathcross}, which is crucial
for the feasibility of our boundary recognition algorithm.
The network nodes know the value of $d$, and the fact that $G$ is a
$d$-QUDG. The embedding $p$ itself is not available to them.

An important property of our algorithms is that they do not require a
specific distribution of the nodes. We only assume the existence of
the embedding $p$.


\section{Boundary Recognition}
\label{sec:bounds}

This section introduces algorithms that detect the boundary of the
region that is covered by the sensor nodes. First, we present some
properties of QUDGs. These allow deriving geometric knowledge from
the network graph without knowing the embedding $p$. Then we
define the Boundary Detection Problem, in which solutions are geometric
descriptions of the network arrangement.  Finally, we describe a start
procedure and an augmentation procedure.  Together, they form a
local improvement algorithm for boundary detection.

\subsection{QUDG Properties.}

We start this section with a simple property of QUDGs. The special
case where $d=1$ was originally proven by Breu and Kirkpatrick
\cite{breu98unit}. Recall that we assume $d\geq \sqrt{2}/2$.

\begin{lemma}\label{thm:pathcross}
  Let $u,v,w,x$ be four different nodes in $V$, where $uv\in E$ and
  $wx\in E$. Assume the straight-line embeddings of $uv$ and $wx$
  intersect. Then at least one of the edges in $F:=\{uw, ux, vw, vx\}$
  is also in $E$.
\end{lemma}

\begin{proof}
  We assume $p(u)\neq p(v)$; otherwise the lemma is trivial.
  Let $a:=\|p(u)-p(v)\|_2\leq 1$. Consider two circles of common
  radius $d$ with their centers at $p(u)$, resp.~$p(v)$. The distance
  between the two intersection points of these circles is
  $h:=2\sqrt{d^2-\tfrac{1}{4}a^2}\geq 1$. If $F$ and $E$ were
  distinct, $p(w)$ and $p(x)$ had both to be outside the two circles.
  Because of the intersecting edge embeddings, $\|p(w)-p(x)\|_2>h\geq
  1$, which would contradict $wx\in E$.
\end{proof}

Lemma~\ref{thm:pathcross} allows to use edges in the graph to separate
nodes in the embedding $p$, even without knowing $p$. We can use this
fact to certify that a node is inside the geometric structure defined
by some other nodes. Let $C\subset V$ be a chordless cycle in $G$,
i.e., $(C,E(C))$ is a connected 2-regular subgraph of $G$. $P(C)$
denotes the polygon with a vertex at each $p(v), v\in C$ and an edge
between vertices whose corresponding nodes are adjacent in $G$. $P(C)$
also defines a decomposition of the plane into faces. A point in the
infinite face is said to be {\em outside} of $P(C)$, all other points
are {\em inside}.

\begin{corollary}\label{thm:pathwitness}
  Let $C$ be a chordless cycle in $G$, and let $U\subset V$ be
  connected. Also assume $N(C)\cap U=\varnothing$. Then either the
  nodes in $U$ are all on the outside of $P(C)$, or all on the inside.
\end{corollary}

This follows directly from Lemma~\ref{thm:pathcross}. So we can use
chordless cycles for defining cuts that separate inside from outside
nodes. Our objective is to certify that a given node set is inside the
cycle, thereby providing insight into the network's geometry.
Unfortunately, this is not trivial; however, it is possible to guarantee 
that a node set is outside the cycle.

Note that simply using two node sets that are separated by a
chordless cycle $C$ and proving that the first set is outside the
cycle does not guarantee that the second set is on the inside. The two
sets could be on different sides of $P(C)$. So we need more
complex arguments to certify insideness.

Now we present a certificate for being on the outside. Define
$\fit_d(n)$ to be the maximum number of independent nodes $J$ that can
be placed inside a chordless cycle $C$ of at most $n$ nodes in any
$d$-QUDG embedding such that $J\cap N(C)=\varnothing$. We say that
nodes are independent, if there is no edge between any two of them.
These numbers exist because independent nodes are placed at least $d$
from each other, so there is a certain area needed to contain the
nodes. On the other hand, $C$ defines a polygon of perimeter at most
$|C|$, which cannot enclose arbitrarily large areas. Also define
$\enc_d(m):=\min\{ n: \fit_d(n)\geq m \}$, the minimum length needed
to fit $m$ nodes.

\begin{table*}
  \centering
  \begin{tabular}{|l|rrrrrrrrrrrrrrrrrrrr|}\hline
    $n$         & 1& 2& 3& 4& 5& 6& 7& 8& 9&10&11&12&13&14&15&16&17&18&19&20\\\hline
    $\fit_1(n)$ & 0& 0& 0& 0& 0& 0& 1& 1& 2& 3& 4& 5& 7& 8& 9&12&14&16&17&19\\\hline
    $\lim_{d\uparrow 1}\fit_d(n)$
    & 0& 0& 0& 0& 0& 1& 1& 2& 3& 4& 5& 7& 8& 9&12&14&16&17&19&23\\\hline
  \end{tabular}
  \caption{First values of $\fit_d(n)$}
  \label{tab:fit}
\end{table*}
The first 20 values of $\fit_1$ and $\fit_{1-\epsilon}$ for some small
$\epsilon$ are shown in Table~\ref{tab:fit}. They can be obtained by
considering hexagonal circle packings. Because these are constants
it is reasonable to assume that the first few values of $\fit_d$ are
available to every node.

We are not aware of the exact values of $\fit_d$ for all $d$. However,
our algorithms just need upper bounds for $\fit_d$, and lower bounds
for $\enc_d$. (An implementation of the following algorithms has to be
slightly adjusted to use bounds instead of exact values.)

%
Now we can give a simple criterion to decide that a node set is
outside a chordless cycle:
\begin{lemma}\label{thm:setisoutside}
  Let $C$ be a chordless cycle and $I\subset V\setminus N(C)$ be a
  connected set that contains an independent subset $J\subset I$. If
  $|J|>\fit_d(|C|)$, then every node in $I$ is outside $P(C)$.
\end{lemma}
\begin{proof}
  By Corollary~\ref{thm:pathwitness} and the definition of $\fit_d$.
\end{proof}

\subsection{Problem statement.}

In this section, we define the Boundary Detection Problem.
Essentially, we are looking for node sets and chordless cycles, where
the former are guaranteed to be on the inside of the latter. For the
node sets to become large, the cycles have to follow the perimeter of
the network region. In addition, we do not want holes in the network
region on the inside of the cycles, to ensure that each boundary is
actually reflected by some cycle.

We now give formal definitions for these concepts. We begin with the
definition of a hole: The graph $G$ and its straight-line embedding
w.r.t.~$p$ defines a decomposition of the plane into faces. A finite
face $F$ of this decomposition is called {\em $h$-hole} with parameter
$h$ if the boundary length of the convex hull of $F$ strictly exceeds
$h$. An important property of an $h$-hole $F$ is the following fact:
Let $C$ be a chordless cycle with $|C|\leq h$. Then all points $f\in
F$ are on the outside of $P(C)$.

To describe a region in the plane, we use chordless cycles in the
graph that follow the perimeter of the region. There is always one
cycle for the outer perimeter. If the region has holes, there is an
additional cycle for each of them. We formalize this in the opposite
direction: Given the cycles, we define the region that is enclosed by
them.  So let $\calC:=(C_b)_{b\in\calB}$ be a family of chordless
cycles in the network. It describes the boundary of the region
$A(\calC)\subset\plane$, which is defined as follows.  First let
$\tilde{A}$ be the set of all points $x\in\plane$ for which the
cardinality of $\{b\in\calB: x\mbox{ is on the inside of }P(C_b)\}$ is
odd. This set gives the inner points of the region, which are all
points that are surrounded by an odd number of boundaries. The
resulting region is defined by
\begin{equation}
  A(\calC) :=
  \bigcup_{b\in\calB}P(C_b) 
  \cup \tilde{A}
  \;.
\end{equation}
\begin{figure}
  \centering
  \setlength\unitlength{1cm}
  \begin{picture}(5,2.7)
    \put(0,.3){\includegraphics[height=2\unitlength]{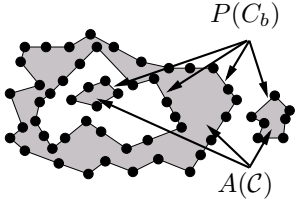}}
    \put(2.7,0){\makebox(1,.4)[t]{$A(\calC)$}}
    \put(2.7,2.3){\makebox(1,.4)[b]{$P(C_b)$}}
  \end{picture}
  \caption{Area described by four boundary cycles.}
  \label{fig:cyclesarea}
\end{figure}
See Figure~\ref{fig:cyclesarea} for an example with some cycles and
the corresponding region.

We can use this approach to introduce geometry descriptions. These consist
of some boundary cycles $(C_b)_{b\in\calB}$, and nodes sets
$(I_i)_{i\in\calI}$ that are known to reside within the described
region. The sets are used instead of direct representations of
$A(\calC)$, because we seek descriptions that are completely
independent of the actual embedding of the network.  There is a
constant $K$ that limits the size of holes. We need $K$ to be large
enough to find cycles in the graph that cannot contain $K$-holes.
Values $K\approx 15$ fulfill these needs.
\begin{Definition}
  A {\em feasible geometry description} (FGD) is a pair
  $(\calC,(I_i)_{i\in\calI})$ with $\calC=(C_b)_{b\in\calB}$ of node set families that
  fulfills the following conditions:\\
(F1) Each $C_b$ is a chordless cycle in $G$ that does not
    contain any node from the family $(I_i)_{i\in\calI}$.\\
(F2) There is no edge between different cycles.\\
(F3) For each $v\in I_i$ ($i\in\calI$), $p(v)\in
    A(\calC)$.\\
(F4) For every component $A'$ of $A(\calC)$,
    there is an index $i\in\calI$, such that $p(v)\in A'$ $\forall v\in
    I_i$ and $p(v)\notin A'$ $\forall v\in I_j, j\neq i$.\\
(F5) $A(\calC)$ does not contain an inner point
    of any $k$-hole for $k>K$.
\end{Definition}
Note that condition (F4) correlates some cycles with a component of
$A(\calC)$, which in turn can be identified by an index $i\in\calI$.
This index is denoted by $\IC(v)$, where $v\in V$ is part of such a
cycle or the corresponding $I_i$.

See Figure~\ref{fig:cex:network} in the computational experience
section for an example network. Figures~\ref{fig:cex:flowers}
and~\ref{fig:cex:augment} show different FGDs in this network.

We are looking for a FGD that has as many inside nodes as possible,
because that forces the boundary cycles to follow the network boundary
as closely as possible. The optimization problem we consider for
boundary recognition is therefore

\begin{equation}
  \mbox{(BD)}\left\{
    \begin{array}{ll}
      \max        & |\cup_{i\in\calI}I_i| \\
      \mbox{s.t.} & ((C_b)_{b\in\calB},(I_i)_{i\in\calI}) \mbox{ is a FGD}
    \end{array}
    \right.
    \;.
\end{equation}

\subsection{Algorithm.}
\label{sec:bounds:algo}

We solve (BD) with local improvement methods that switch from one FGD
to another of larger $|\cup_{i\in\calI}I_i|$. In addition to the FGD,
our algorithms maintain the following sets:
\begin{itemize}
\item The set $C:=\cup_{b\in\calB}C_b$ of {\em cycle nodes}.
\item $N(C)$, the {\em cycle neighbors}. Notice $C\subseteq N(C)$.
\item $I:=\cup_{i\in\calI}I_i$, the {\em inner nodes}. Our algorithms
  ensure $I\cap N(C)=\varnothing$ (this is no FGD requirement), and
  all $I_i$ will be connected sets. This is needed in several places
  for Lemma~\ref{thm:setisoutside} to be applicable.
\item $J\subseteq I$, consisting of so-called {\em independent
    inners}. These nodes form an independent set in $G$.
\item the set $U:=V\setminus(N(C)\cup I)$ of {\em unexplored} nodes.
\end{itemize}

Initially, $U=V$, all other sets are empty.

We need to know how many independent nodes are in a given $I_i$ as
proof that a cycle cannot accidently surround $I_i$. Because all
considered cycles consist of at most $K$ nodes, every count exceeding
$\fit_d(K)$ has the same implications. So we measure the mass of an
$I_i$ by
\begin{equation}
  M(i) := \min\{ |J\cap I_i|, \fit_d(K)+1 \}\;.
\end{equation}

Because we are interested in distributed algorithms, we have to
consider what information is available at the individual nodes. Our
methods ensure (and require) that each node knows to which of the
above sets it belongs. In addition, each cycle node $v\in C$ knows
$\IC(v)$, $M(\IC(v))$, and $N(v)\cap C$, and each cycle neighbor $w\in
N(C)$ knows $N(w)\cap C$.

The two procedures are described in the following two sections: First
is an algorithm that produces start solutions, second an
augmentation method that increases the number of inside nodes.

\subsection{Flowers.}

\begin{figure} 
  \centering
  \begin{minipage}[t]{3.7cm}
    \centering
    \setlength{\unitlength}{.74cm}
    \newcommand{\lb}[1]{{\footnotesize$#1$}}
    \newcommand{\fij}[2]{\lb{f_{\!#1\makebox[.08cm]{,}#2}}}
    \begin{picture}(5,5)
      \put(0,0){\includegraphics[height=5\unitlength]{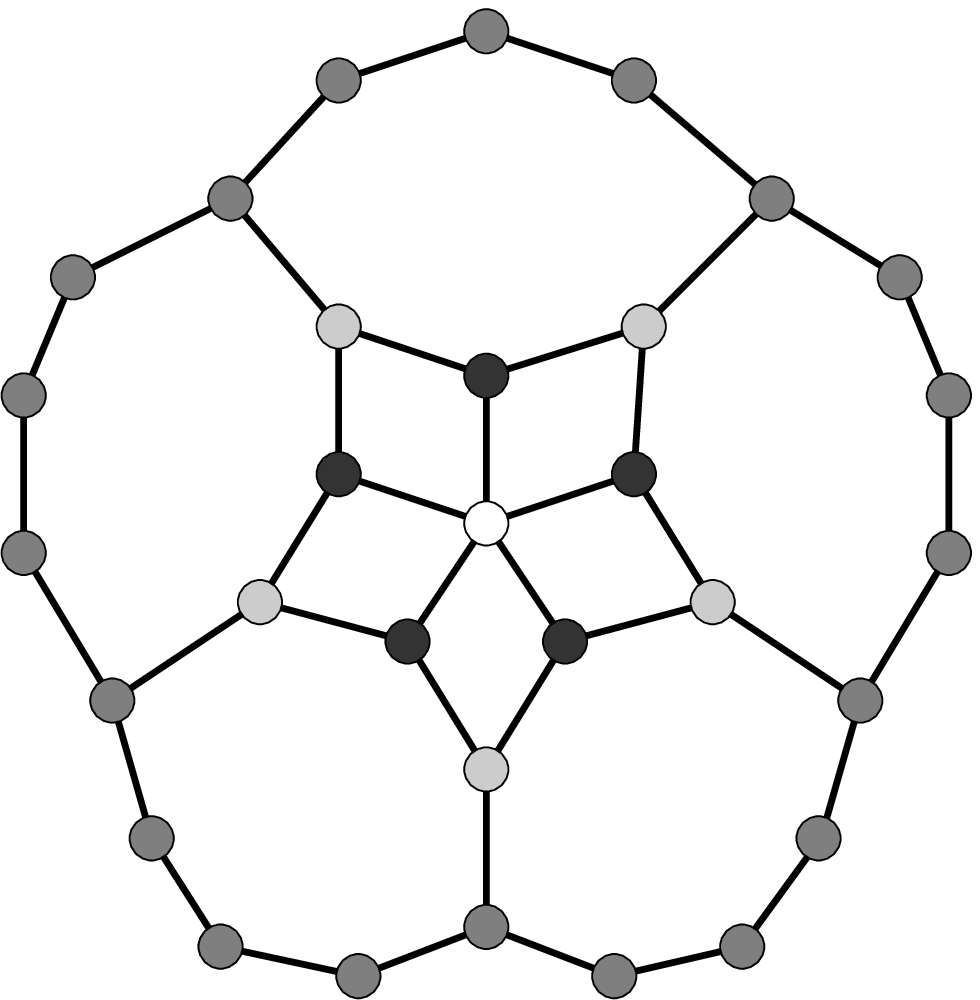}}
      \put(2.4,2.6){\lb{f_{\!0}}}
      \put(3.3,2.6){\fij{1}{1}}
      \put(3.3,3.2){\fij{2}{1}}
      \put(2.1,3.4){\fij{1}{5}}
      \put(3.7,2.0){\fij{2}{2}}
      \put(2.8,1.5){\fij{1}{2}}
      \put(2.5,1.0){\fij{2}{3}}
      \put(1.5,1.5){\fij{1}{3}}
      \put( .8,2.2){\fij{2}{4}}
      \put(1.0,2.6){\fij{1}{4}}
      \put(1.0,3.3){\fij{2}{5}}
      \put( .6,4.2){\fij{3}{5}}
      \put(3.8,4.2){\fij{3}{1}}
      \put(4.4,1.4){\fij{3}{2}}
      \put(2.1, .0){\fij{3}{3}}
      \put( .0,1.2){\fij{3}{4}}
      \put(4.2,2.5){\lb{W_1}}
      \put(3.3, .6){\lb{W_2}}
      \put(1.1, .5){\lb{W_3}}
      \put( .2,2.5){\lb{W_4}}
      \put(2.2,4.3){\lb{W_5}}
    \end{picture}
    \caption{A 5-flower.}
    \label{fig:flower}
  \end{minipage}
  \hfill
  \begin{minipage}[t]{4.6cm}
    \centering
    \setlength{\unitlength}{0.4cm}
    \begin{picture}(11.5,9.25)
      \put(0,.2){(same X coordinates as Y)}
      \put(0,1){
        \put(0,.5){\includegraphics[height=6.5\unitlength]{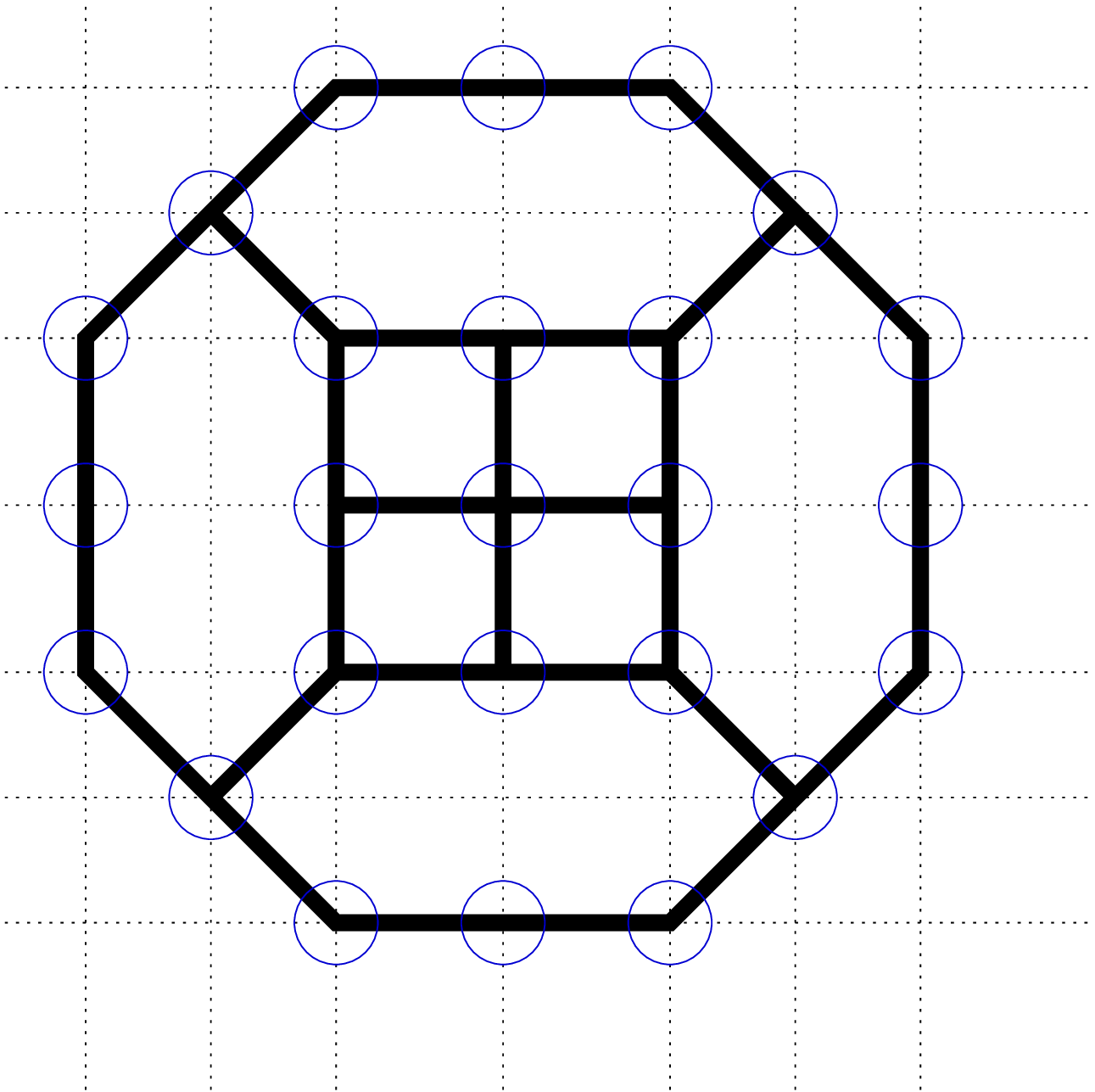}}
        \put(.2,-.18){
          \put(6.5,4){{\footnotesize $0$}}
          \put(6.5,5){{\footnotesize $+w$}}
          \put(6.5,3){{\footnotesize $-w$}}
          \put(6.5,2.3){{\footnotesize $-(1+\sqrt{2}/2)w$}}
          \put(6.5,1.5){{\footnotesize $-(1+\sqrt{2})w$}}
        }
        \put(6.7,5.6){{\footnotesize $(1+\sqrt{2}/2)w$}}
        \put(6.7,6.4){{\footnotesize $(1+\sqrt{2})w$}}
      }
    \end{picture}
    \caption{Construction of a 4-flower in a dense region.}
    \label{fig:4flowercons}
  \end{minipage}
\end{figure}

So far, we have presented criteria by which one can decide that some nodes
are {\em outside} a chordless cycle, based on a packing argument. Such
a criterion will not work for the {\em inside}, as any set of nodes that fit
in the inside can also be accomodated by the unbounded outside.
Instead, we now present a stronger strcutural criterion that is based
on a particular subgraph, an $m$-flower. For such a structure, we can prove
that there are some nodes on the inside of a chordless cycle. Our
methods start by searching for flowers, leading to a FGD.
We begin by actually defining a flower, see
Figure~\ref{fig:flower} for a visualization.
\begin{Definition}
  An $m$-flower in $G$ is an induced subgraph whose node set consists
  of a seed $f_0\inV$, independent nodes $f_{1,1},\ldots,f_{1,m}\inV$,
  bridges $f_{2,1},\ldots,f_{2,m}\inV$, hooks
  $f_{3,1},\ldots,f_{3,m}\inV$, and chordless paths $W_1,\ldots,W_m$, where each
  $W_i=(w_{j,1},\ldots,w_{j,\ell_j})\subset V$. All of these
  $1+3m+\sum_{j=1}^m\ell_j$ nodes have to be different nodes. For
  convenience, we define $f_{j,0}:=f_{j,m}$ and $f_{j,m+1}:=f_{j,1}$
  for $j=1,2,3$.

  The edges of the subgraph are the following: The seed $f_0$ is
  adjacent to all independent nodes: $f_0f_{1,j}\in E$ for
  $j=1,\ldots,m$.  Each independent node $f_{1,j}$ is connected to two
  bridges: $f_{1,j}f_{2,j}\in E$ and $f_{1,j}f_{2,j+1}\in E$.  The
  bridges connect to the hooks: $f_{2,j}f_{3,j}\in E$ for
  $j=1,\ldots,m$.  Each path $W_j$ connects two hooks, that is,
  $f_{3,j}w_{j,1}, w_{j,1}w_{j,2},\ldots,w_{j,\ell_j}f_{3,j+1}$ are
  edges in $E$.
  
  Finally, the path lengths $\ell_j$, $j=1,\ldots,m$ obey
  \begin{eqnarray}
    \fit_d(5+\ell_j) &<& m-2 
    \;,
    \label{eq:flower-emptypetal}\\
    \fit_d(7+\ell_j) &<&
    \left\lceil \frac{1}{2}\left(\sum_{k\neq i}\ell_k +1\right)\right\rceil
    \;.
    \label{eq:flower-outerpetal}
  \end{eqnarray}
\end{Definition}

Notice that Equations~\eqref{eq:flower-emptypetal}
and~\eqref{eq:flower-outerpetal} can be fulfilled: for $d=1$, $m=5$
and $\ell_1=\ell_2=\ldots=\ell_5=3$ are feasible. This is the flower
shown in Figure~\ref{fig:flower}.

The beauty of flowers lies in the following fact:
\begin{lemma}
  In every $d$-QUDG embedding of a $m$-flower, the independent nodes
  are placed on the inside of $P(C)$, where
  $C:=\{f_{3,1},\ldots,f_{3,m}\}\cup\bigcup_{j=1}^mW_j$ is a chordless
  cycle.
\end{lemma}

\begin{proof}
  Let $P_j:=(f_{1,j},f_{2,j},f_{3,j},W_j,f_{3,j+1},f_{2,j+1})$ be a
  petal of the flower. $P_j$ defines a cycle of length $5+\ell_j$. The
  other nodes of the flower are connected and contain $m-2$
  independent bridges. According to~\eqref{eq:flower-emptypetal}, this
  structure is on the outside of $P(P_j)$.

  Therefore, the petals form a ring of connected cycles, with the seed
  on either the inside or the outside of the structure. Assume the
  seed is on the outside. Consider the infinite face of the
  straight-line embedding of the flower. The seed is part of the outer
  cycle, which consists of $7+\ell_j$ nodes for some
  $j\in\{1,\ldots,m\}$. This cycle has to contain the remaining flower
  nodes, which contradicts ~\eqref{eq:flower-outerpetal}. Therefore,
  the seed is on the inside, and the claim follows.
\end{proof}

Because we do not assume a particular distribution of the nodes, we
cannot be sure that there is a flower in the network. Intuitively,
this is quite clear, as any node may be close to the boundary, so that
there are no interior nodes; as the nodes can only make use of the
local graph structure, and have no direct way of detecting region
boundaries, this means that for low densities everywhere, our
criterion may fail.  As we show in the following, we can show the
existence of a flower if there is a densely populated region somewhere:
We say $G$ is {\em locally $\epsilon$-dense} in a region
$A\subset\plane$, if every $\epsilon$-ball in $A$ contains at least
one node, i.e., $\forall z\in\plane: B_{\epsilon}(z)\subset
A\Rightarrow\exists v\in V:\|p(v)-z\|_2\leq\epsilon$.

\begin{lemma}
  Let $0<\epsilon<\tfrac{3}{2}-\sqrt{2}\approx 0.086$. Assume $d=1$.  If $G$ is
  $\epsilon$-dense on the disk $B_3(z)$ for some $z\in\plane$, then
  $G$ contains a $4$-flower.
\end{lemma}

\begin{proof} Let $w:=2(\sqrt{2}-1)$. See
  Figure~\ref{fig:4flowercons}.  Place an $\epsilon$-ball at all the
  indicated places and choose a node in each. Then the induced
  subgraph will contain precisely the drawn edges. Then $m=4$ and
  $\ell_1=\ldots=\ell_4=3$, so for $d=1$, these $\ell$-numbers are
  feasible.
\end{proof}


Now we present the actual algorithm to detect flowers. Notice that a
flower is a strictly local structure, so we use a very simple kind of
algorithm. Each node $v\inV$ performs the following phases after the
simultaneous wakeup:

\indent 1. Collect the subgraph on $N_8(v)$.\\
\indent 2. Find a flower.\\
\indent 3. Announce update.\\
\indent 4. Update.

\paragraph{Collect:} First, each node $v\inV$ collects and stores the
local neighborhood graph $N_8(v)$. This can be done in time
$\calO(\Delta_1)$ and message complexity $\calO(\Delta_1\Delta_8)$, if
every nodes broadcasts its direct neighborhood to its 8-neighborhood.

\paragraph{Find Flower:} Then, every node decides for itself whether it is
the seed of a flower. This does not involve any communication.

\paragraph{Announce update:} Because there could be multiple
intersecting flowers, the final manifestation of flowers has to be
scheduled: Every seed of some flower broadcasts an announcement to all
nodes of the flower and their neighbors. Nodes that receive
multiple announcements decide which seed has higher priority, e.g.,
higher ID number. The seeds are then informed whether they lost such a
tie-break. This procedure has runtime $\calO(1)$ and message
complexity $\calO(\Delta_9)$ per seed, giving a total message
complexity of $\calO(\Delta_9|V|)$.

\paragraph{Update:} The winning seeds now inform their flowers that
the announced updates can take place. This is done in the same manner
as the announcements. The nodes that are part of a flower store their
new status and the additional information described in
Section~\ref{sec:bounds:algo}.

\subsection{Augmenting Cycles.}

Now that we have an algorithm to construct an initial FGD in the
network, we seek an improvement method. For that, we employ {\em
  augmenting cycles}. Consider an FGD
$((C_b)_{b\in\calB},(I_i)_{i\in\calI})$. Let
$U=(u_1,u_2,\ldots,u_{|U|})\subset V$ be a (not necessarily chordless)
cycle. For convenience, define $u_0:=u_{|U|}$ and $u_{|U|+1}:=u_1$.

When augmenting, we open the cycles $(C_b)_{b\in\calB}$ where they
follow $U$, and reconnect the ends according to $U$. Let
$U^-:=\{u_i\in U: u_{i-1},u_i,u_{i+1}\in C\}$ and $U^+:=U\setminus C$.
The resulting cycle nodes of the augmentation operation are then
$C':=C\cup U^+\wo U^-$. If $N(U)\cap I=\varnothing$, this will not
affect inside nodes, and it may open some new space for the inside
nodes to discover. In addition, as the new cycle cannot contain a
$|U|$-hole, we can limit $|U|$ to guarantee condition (F5).

\newcommand{\iniv}{v_1} We use a method that will search for an
augmenting cycle that will lead to another FGD with a larger number of
inside nodes, thereby performing one improvement step. The method 
is described for a single node $\iniv\in C$ that searches for an
augmenting cycle containing itself. This node is called {\em
  initiator} of the search.

It runs in the following phases:

\indent 1. Cycle search.\\
\indent 2. Check solution.\\
\indent\indent (a) Backtrack.\\
\indent\indent (b) Query feasibility.\\
\indent 3. Announce update.\\
\indent 4. Update.

\paragraph{Cycle search:} $\iniv$ initiates the search by passing around a
token. It begins with the token $T=(\iniv)$.  Each node that
receives this token adds itself to the end of it and forwards it to a
neighbor. When the token returns from there, the node forwards it to the
next feasible neighbor. If there are no more neighbors, the node removes
itself from the list end and returns the token to its predecessor.

The feasible neighbors to which $T$ gets forwarded are all nodes in
$V\wo I$. The only node that may appear twice in the token is $\iniv$,
which starts the ``check solution'' phase upon reception of the token.
In addition, $T$ must not contain a cycle node between two cycle
neighbors. The token is limited to contain
\begin{eqnarray}
|T| &<& \min_{v\in T\cap C} \enc_d(M(\IC(v))) \label{eq:tokensizelimit}\\
    &\leq& K \label{eq:tokengenlimit}
\end{eqnarray}
 nodes. This phase can be implemented such that
no node (except for $\iniv$) has to store any information about the
search.
When this phase terminates unsuccessfully, i.e., without an identified
augmenting cycle, the initiator exits the algorithm.

\paragraph{Check Solution:} When the token gets forwarded to $\iniv$,
it describes a cycle. $\iniv$ then sends a backtrack message backwards
along $T$:

\paragraph{Backtrack:} While the token travels backwards, each node
performs the following: If it is a cycle node, it broadcasts a query
containing $T$ to its neighbors, which in turn respond whether they
would become inside nodes after the update. Such nodes are called {\em
  new inners}.  Then, the cycle node stores the number of positive
responses in the token.

A non-cycle node checks whether it would have any chords after the
update. In that case, it cancels the backtrack phase and informs
$\iniv$ to continue the cycle search phase.

\paragraph{Query Feasibility:} When the backtrack message reaches
$\iniv$, feasibility is partially checked by previous steps.  Now,
$\iniv$ checks the remaining conditions.

Let $\calI':=\{\IC(v):v\in C\cap T\}$. First, it confirms that for
every $i\in\calI$ there is a matching cycle node in the token that has
a nonzero new inner count. Then it picks a $i'\in\calI$. All new
inners of cycle nodes of this $\IC$ value then explore the new inner
region that will exist after the update. This can be done by a BFS
that carries the token. The nodes report back to $\iniv$ the $\IC$
values of new inner nodes that could be reached. If this reported set
equals $\calI'$, $T$ is a feasible candidate for an update and phase
``announce update'' begins. Otherwise, the cycle search phase
continues.

\paragraph{Announce update:} Now $T$ contains a feasible augmenting
cycle. $\iniv$ informs all involved nodes that an update is coming up.
These nodes are $T$, $N(T)$ and all nodes that can be reached from any
new inner in the new region. This is done by a distributed BFS as in the
``query feasibility'' phase. Let $I'$ be the set of all nodes that
will become inner nodes after the update. During this step, the set
$J$ of independent nodes is also extended in a simple greedy fashion.

If any node receives multiple update announcements, the initiator node
of higher ID wins. The loser is then informed that its announcement
failed.

\paragraph{Update:} When the announcement successfully reached all
nodes without losing a tie-break somewhere, the update is performed.

If there is just one component involved, i.e., $|\calI'|=1$, the
update can take place immediatly.

If $|\calI'|>1$, there might be problems keeping $M(\IC(\cdot))$
accurate if multiple augmentations happen simultaneously. So $\iniv$
first decides that the new ID of the merged component will be $\iniv$.
It then determines what value $M(\IC(\iniv))$ will take after the
update. If this value strictly exceeds $\fit_d(K)$, $M(\IC(\iniv))$ is
independent of potential other updates; the update can take place
immediately. However, $M(\IC(\iniv))\leq \fit_d(K)$, concurrent
updates have to be schedules. So $\iniv$ floods the involved
components with an update announcement, and performs its update after
all others of higher prioity, i.e., higher initiator ID.

Finally, all nodes in $T$ flood their $\tfrac{K}{2}$-hop neighborhood
so that cycle nodes whose cycle search phase was unsuccessful can
start a new attempt, because their search space has changed.

\begin{lemma}
  If the augmenting cycle algorithm performs an update on a FGD, it
  produces another FGD with strictly more inner nodes.
\end{lemma}

\begin{proof}
  We need to show that all five FGD conditions are met: (F1) and (F2)
  are checked in the backtrack phase, (F3) follows from
  \eqref{eq:tokensizelimit}, (F4) from the connectivity test in the
  feasibility check phase, and (F5) follows from
  \eqref{eq:tokengenlimit}. The increase in inner nodes is assured in
  the query feasibility phase.
\end{proof}

\begin{lemma}
  One iteration of the augmenting cycle algorithm for a given
  initiator nodes has message complexity $\calO(\Delta_K^K|V|)$ and time
  complexity $\calO(\Delta_K^K\Delta_1+|V|)$.
\end{lemma}

\begin{proof}
  There at at most $\Delta_K^K$ cycles that are checked. For one
  cycle, the backtrack phase takes $\calO(\Delta_1)$ message and time complexity.
  The query feasibility phase involves flooding the part of the new
  inside that is contained in the cycle. Because there can be any
  number of nodes in this region, message complexity for this flood is
  $\calO(|V|)$. The flood will be finished after at most $2\fit_d(K)$
  communication rounds, the time complexity is therefore $\calO(1)$.
  After a feasible cycle was found, the announce update and update
  phases happen once. Both involve a constant number of floods over
  the network, their message and time complexities are therefore
  $\calO(|V|)$. Combining these complexities results in the claimed
  values.
\end{proof}


\section{Topological Clustering}
\label{sec:cluster}

This section deals with constructing clusters that follow the
geometric network topology. We use the working boundary detection from
the previous section and add a method for clustering.

\subsection{Problem statement.}

We assume the boundary cycle nodes are numbered, i.e.,
$C_b=(c_{b,1},\ldots,c_{b,|C_b|})$ for $b\in\calB$. We use a measure
$\cdist$ that describes the distance of nodes in the subgraph $(C,E(C))$:
\[
  \cdist(c_{b,j},c_{b',j'}) 
  := 
  \left\{\begin{array}{cl}
      +\infty                     &\!\!\!\mbox{if }b\neq b'\\
      \min\{|j'-j|,|C_b|-|j'-j|\} &\!\!\!\mbox{if }b=b'
    \end{array}\right.
\]
That is, $\cdist$ assigns nodes on the same boundary their distance
within this boundary, and $\infty$ to nodes on different boundaries.

For each node $v\in V$, let $Q_v\in C$ be the set of cycle nodes that
have minimal hop-distance to $v$, and let $s_v$ be this distance.
These nodes are called {\em anchors} of $v$. Let $v\in V$ and $u,w\in
N(v)$. We say $u$ and $w$ have {\em distant} anchors w.r.t.~$v$, if
there are $q_u\in Q_u$ and $q_v\in Q_v$ such that $\cdist(q_u,q_w) >
\pi(s_v+1)$ holds (with $\pi=3.14\ldots$). This
generalizes closeness to multiple boundaries to the
closeness to two separate pieces of the same boundary. (Here
``separate pieces'' means that there is sufficient distance
along the boundary between the nodes to form a
half-circle around $v$.)

$v$ is called $k$-{\em Voronoi} node, if $N(v)$ contains at least $k$
nodes with pairwise distant anchors. We use these nodes to identify
nodes that are precisely in the middle between some boundaries.  Let
$V_k$ be the set of all $k$-Voronoi nodes. Our methods are based on
the observation that $V_2$ forms strips that run between two
boundaries, and $V_3$ contains nodes where these strips meet.

The connected components of $V_3$ are called {\em intersection cores}.
We build {\em intersection clusters} around them that extend to the
boundary. The remaining strips are the base for {\em street clusters}
connecting the intersections.

\subsection{Algorithms.}

We use the following algorithm for the clustering:\\
\indent 1. Synchronize end of boundary detection.\\
\indent 2. Label boundaries.\\
\indent 3. Identify intersection cores.\\
\indent 4. Cluster intersections and streets.

\paragraph{Synchronize:} The second phase needs to be started at all
cycle nodes simultaneously, after the boundary detection terminates.
For that matter, we use a synchronization tree in the network, i.e., a
spanning tree. Every node in the tree keeps track of whether there are
any active initiator nodes in their subtree.  When the synchronization
tree root detects that there are no more initiators, it informs the
cycle nodes to start the second phase.  Because the root knows the
tree depth, it can ensure the second phase starts in sync.

\paragraph{Label:} Now the cycle nodes assign themselves consecutive
numbers. Within each cycle $C_b$, this starts at the initiator node of
the last augmentation step. If $C_b$ stems from a flower that has not
been augmented, some node that has been chosen by the flower's seed
takes this role. This start node becomes $c_{b,1}$. It then sends a
message around the cycle so that each node knows its position. Afterwards,
it sends another message with the total number of nodes in the cycle.
In the end, each node $c_{b,j}$ knows $b$, $j$, and $|C_b|$. Finally,
the root of the synchronization tree gets informed about the
completion of this phase.

\paragraph{Intersection cores:} This phase identifies the intersection
cores. It starts simultaneously at all cycle nodes. This is scheduled
via the synchronization tree. This tree's root knows the tree depth.
Therefore, it can define a start time for this phase and broadcast a
message over the tree that reaches all nodes shortly before this time.
Then the cycle nodes start a BFS so that every node $v$ knows one
$q_v\in Q_v$ and $s_v$. The BFS carries information about the anchors
so that $v$ also knows $b$ and $j$ for which $q_v=c_{b,j}$. Also, each
nodes stores this information for all of its neighbors.

Each node $v$ checks whether there are three nodes $u_1,u_2,u_3\in
N(v)$ whose known anchors are distant, i.e., $\cdist(q_{u_j},q_{u_k})
> \pi(s_v+1)$ for $j\neq k$. In that case, $v$ declares itself to be a
3-Voronoi node. This constructs a set $\tilde{V}_3\subseteq V_3$.

Finally, the nodes in $\tilde{V}_3$ determine their connected
components and the maximal value of $s_v$ within each component by
constructing a tree within each component, and assign each component
an unique ID number.

\paragraph{Cluster:} Now each intersection core starts BFS up to the
chosen depth. Each node receiving a BFS message associates with the
closest intersection core. This constructs the intersection clusters.
Afterwards, the remaining nodes determine their connected components
by constructing a tree within each component, thereby forming street
clusters.

Because the synchronization phase runs in parallel to the boundary
detection algorithm, it makes sense to analyze the runtime behaviour
of this phase separately:
\begin{theorem}
  The synchronization phase of the algorithm has both message and time
  complexity $\calO(|V|^3)$.
\end{theorem}

\begin{proof} We do not separate between time and message complexity,
  because here they are the same.  Constructing the tree takes
  $\calO(|V|\log|V|)$, and the final flood is linear. However, keeping
  track of the initiators is more complex: There can be $\calO(|V|)$
  augmentation steps. In each step, $\calO(|V|)$ may change their
  status, which has to be broadcast over $\calO(|V|)$ nodes.
\end{proof}

\begin{theorem}
  The remaining phases have message and time complexity
  $\calO(|V|\log|V|)$.
\end{theorem}

\begin{proof}
  The most expensive operation in any of the phases is a BFS over the
  whole network, which takes $\calO(|V|\log|V|)$.
\end{proof}

%
%
%
%


\section{Computational Experience}

\begin{figure}
  \newlength{\xthishoho}\setlength{\xthishoho}{4.1cm}
  \centering
  \begin{minipage}[t]{\xthishoho}
    \includegraphics[width=\xthishoho]{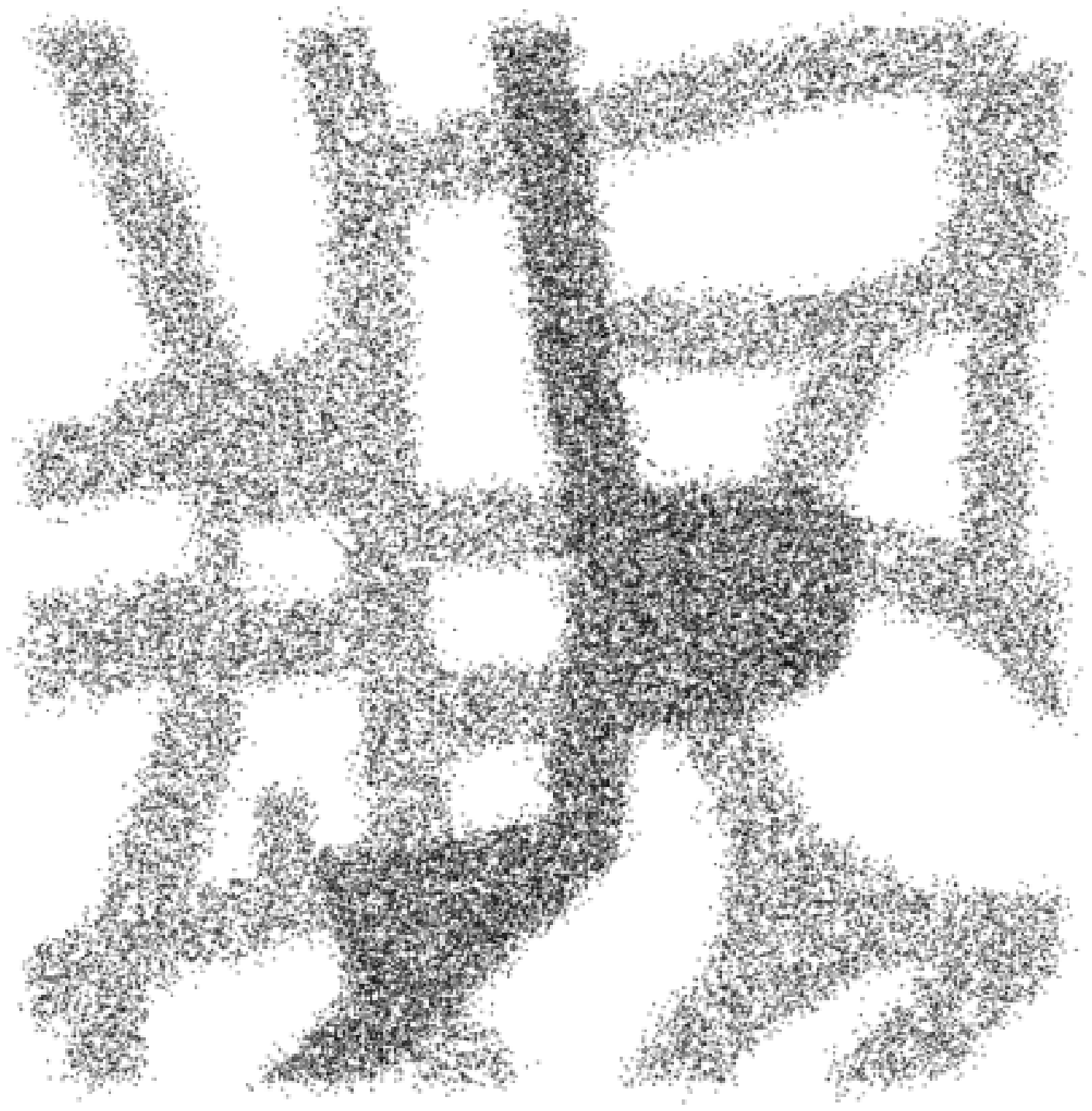}
    \caption{Example network.}
    \label{fig:cex:network}
  \end{minipage}
  \hfill
  \begin{minipage}[t]{\xthishoho}
  \centering
  \includegraphics[width=\xthishoho]{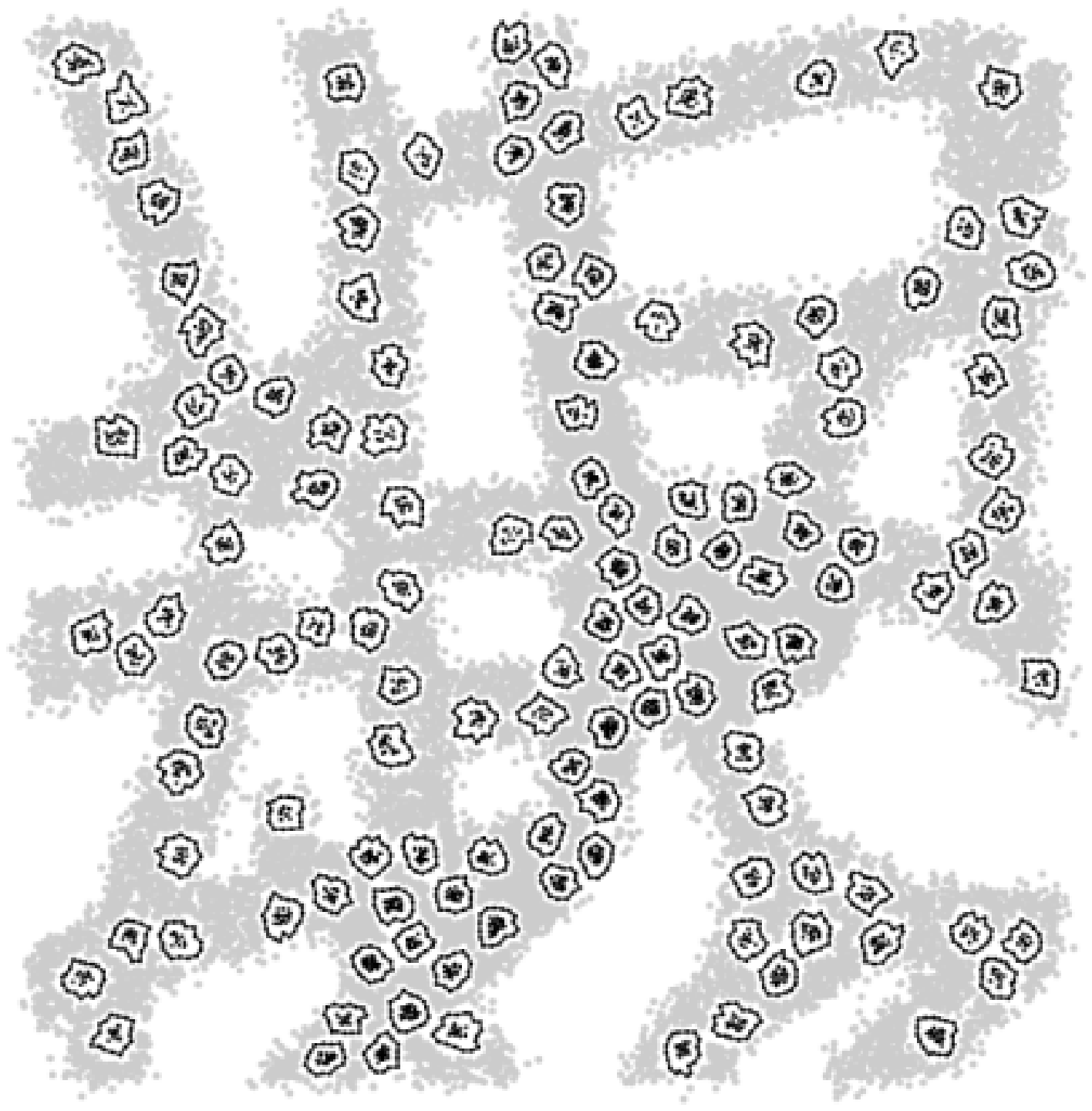}
  \caption{Boundary cycles and inside nodes identified by the
    flower procedure.}
  \label{fig:cex:flowers}
  \end{minipage}
\end{figure}

We have implemented and tested our methods with our large-scale
network simulator {\sc Shawn}~\cite{kpbff-snaswsn-05}.  We demonstrate
the performance on a complex scenario, shown in
Figure~\ref{fig:cex:network}: The network consists of 60,000 nodes
that are scattered over a street map. To show that the procedures do
not require a nice distribution, we included fuzzy boundaries and
varying densities.  Notice that this network is in fact very sparse:
The average neighborhood size is approximatly 20 in the lightly
populated and 30 in the heavily populated area.

\begin{figure}
  \centering
  \includegraphics[height=2.6cm]{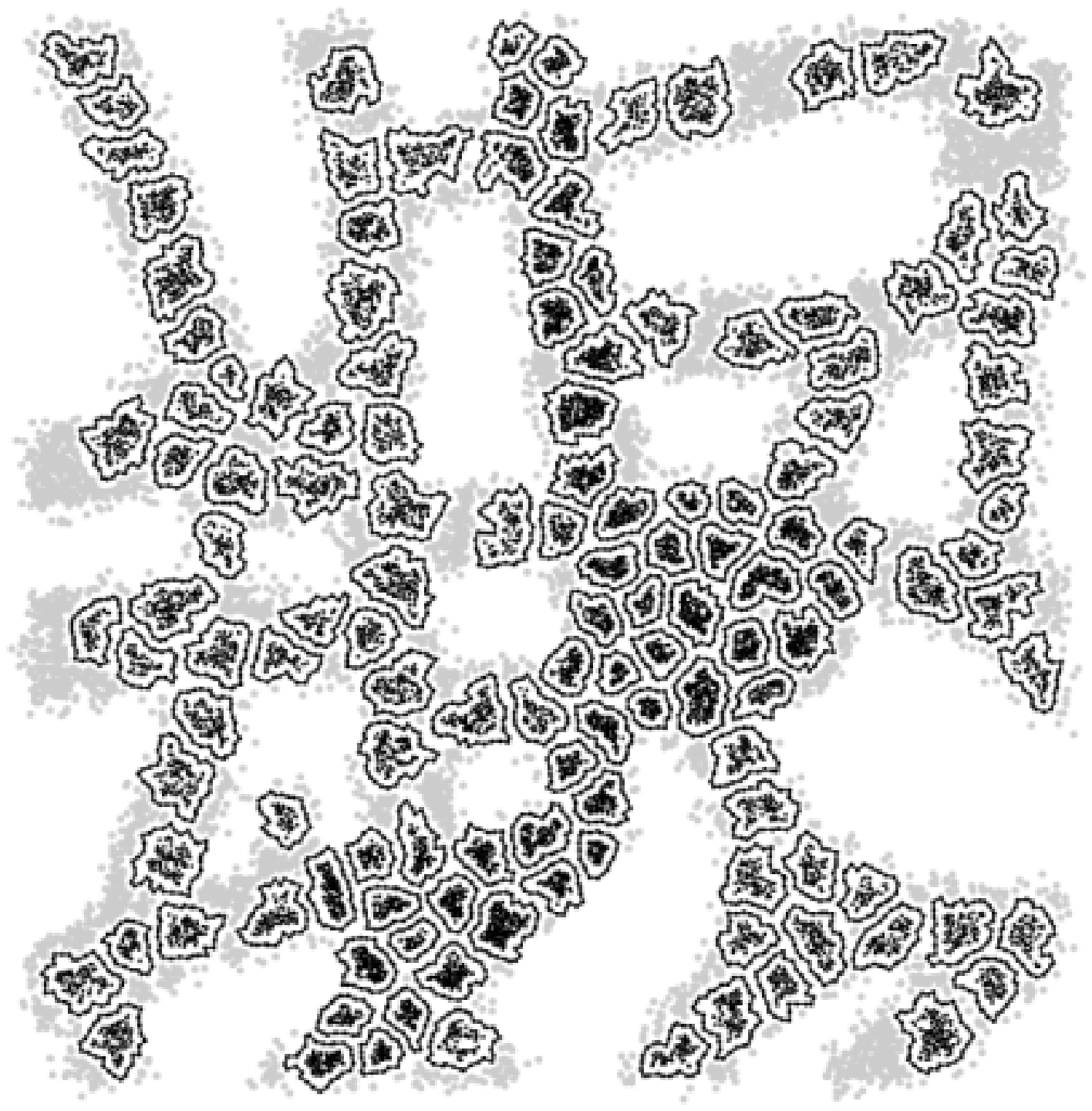}
  \includegraphics[height=2.6cm]{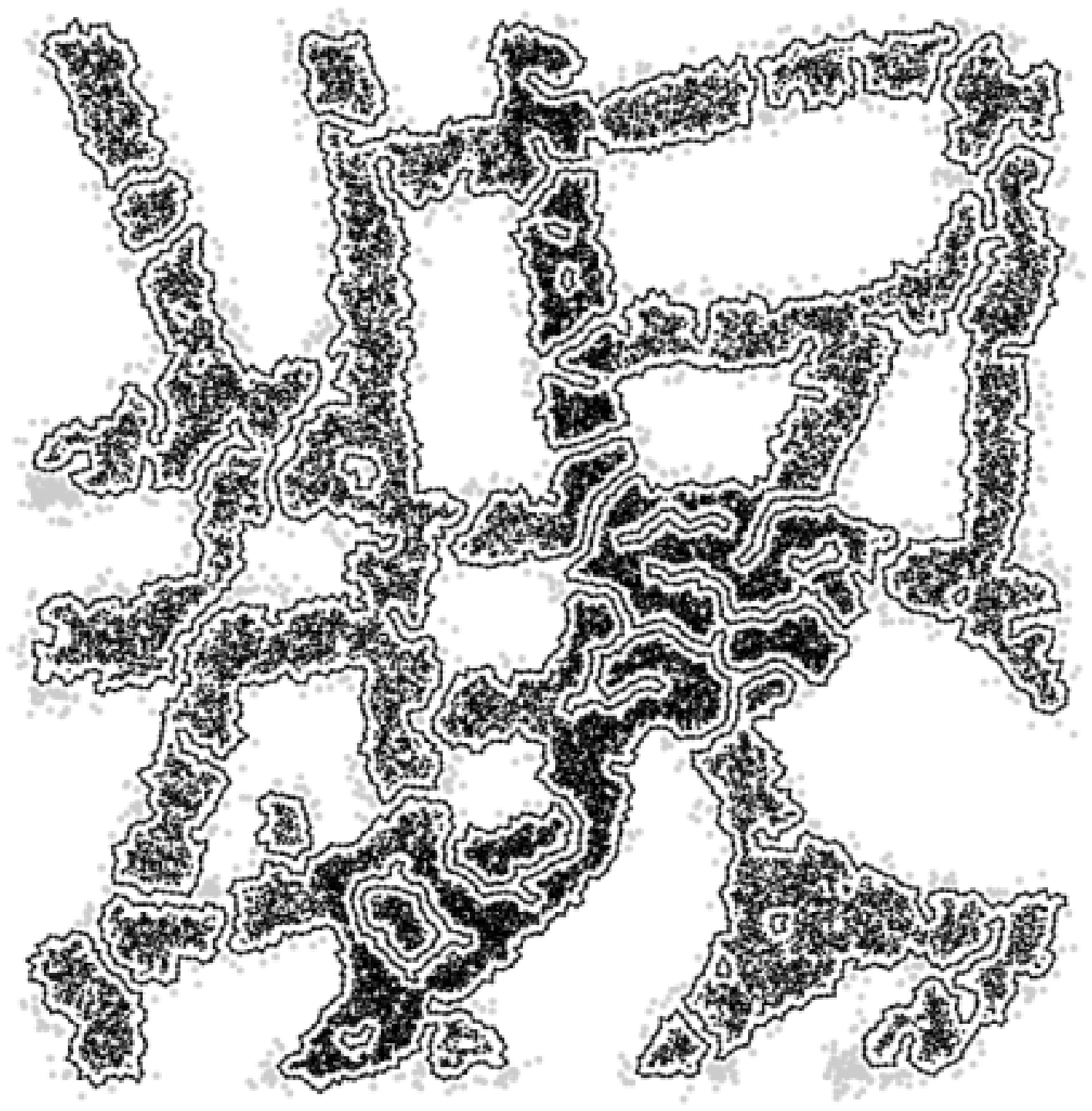}
  \includegraphics[height=2.6cm]{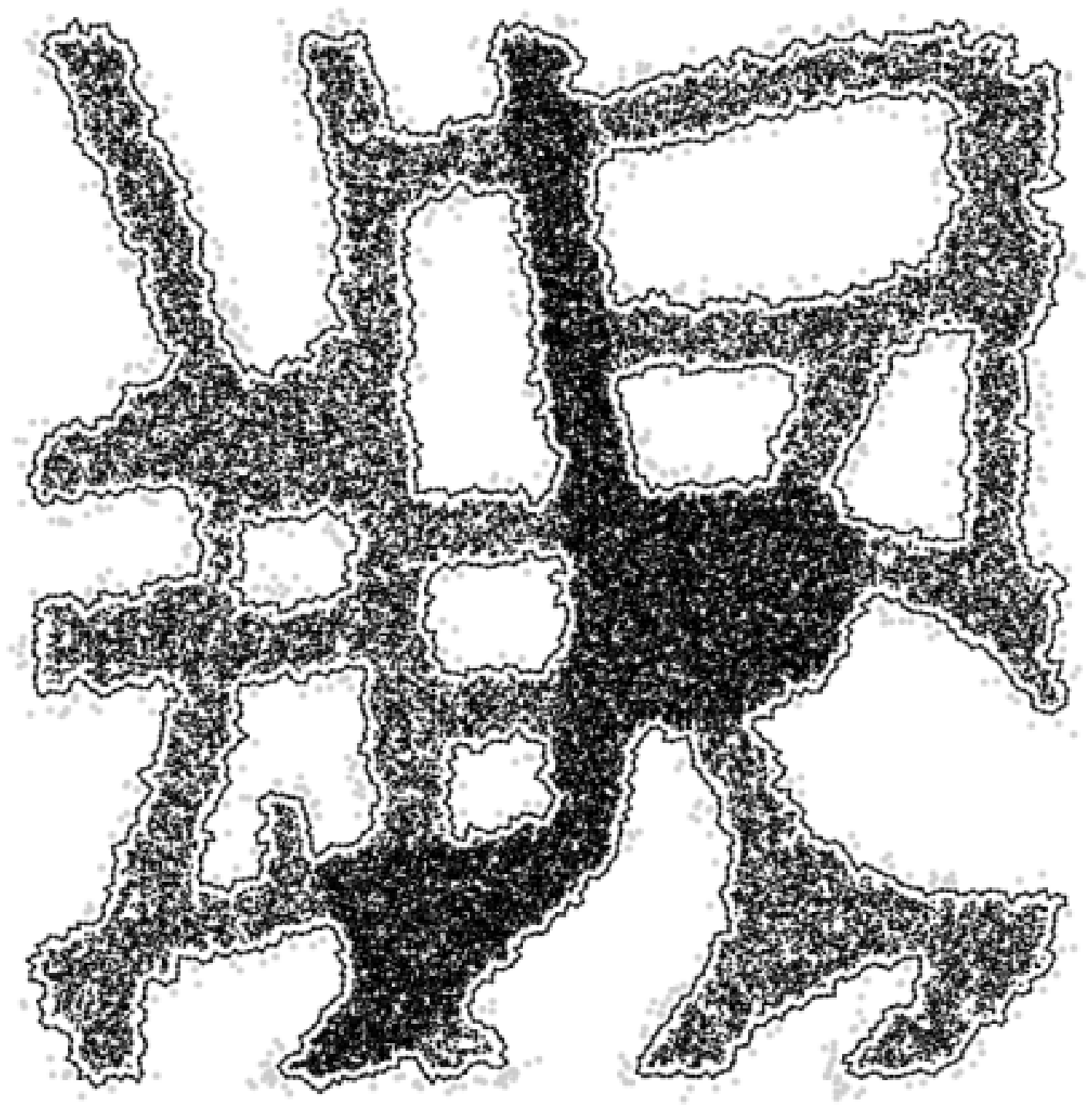}
  \caption{Two snapshots and final state of the
    Augmenting Cycle algorithm.}
  \label{fig:cex:augment}
\end{figure}

Figure~\ref{fig:cex:flowers} shows the FGD that is produced by the
flower procedure. It includes about 70 flowers, where a single one
would suffice to start the augmentation.  Figure~\ref{fig:cex:augment}
shows some snapshots of the augmenting cycle method and its final
state. In the beginning, many extensions to single cycles lead to
growing zones. In the end, they get merged together by multi-cycle
augmentations.  It is obvious that the final state indeed consists of
a FGD that describes the real network boundaries well.

\begin{figure}
  \centering
  \includegraphics[height=4cm]{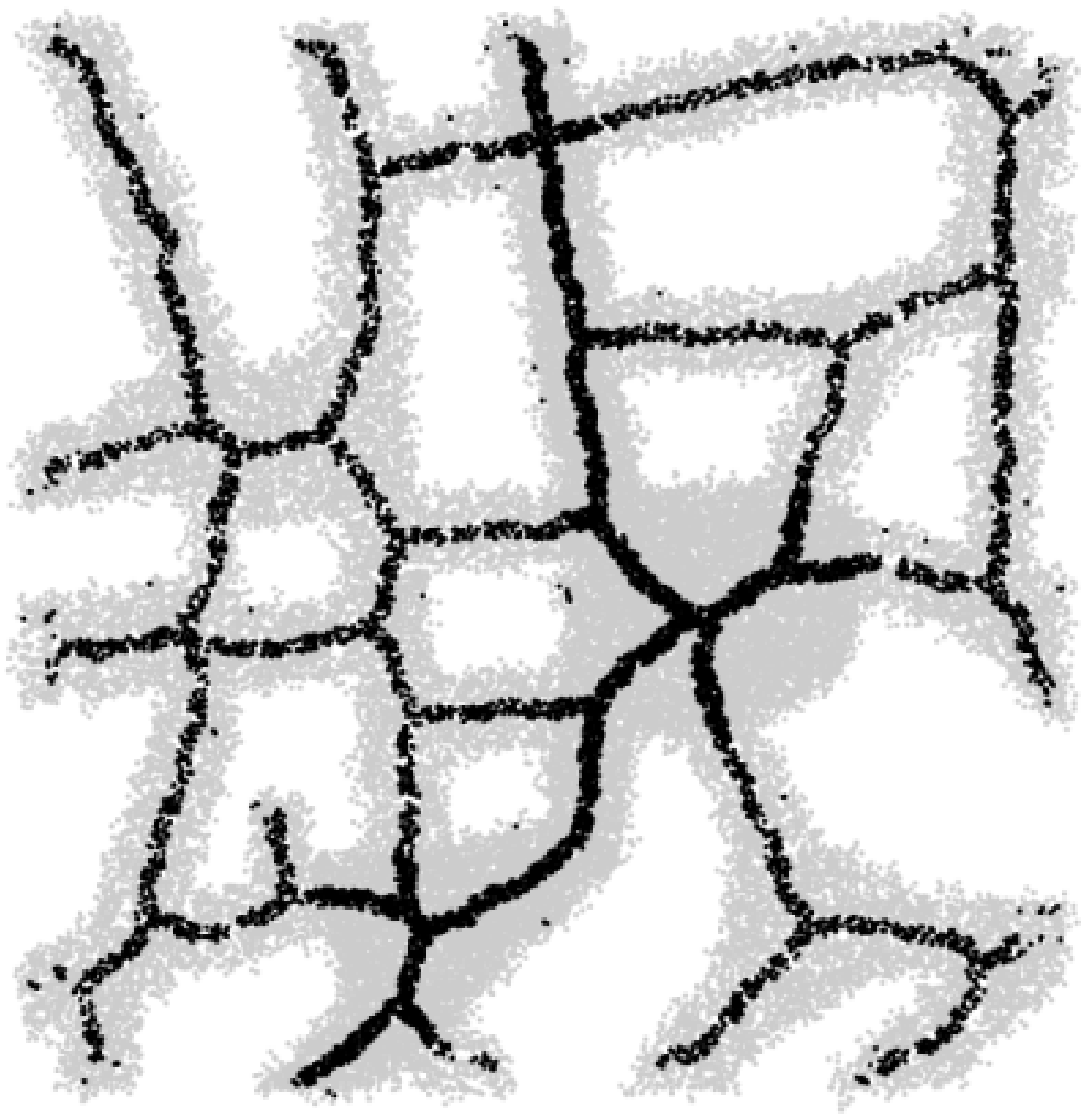}
  \includegraphics[height=4cm]{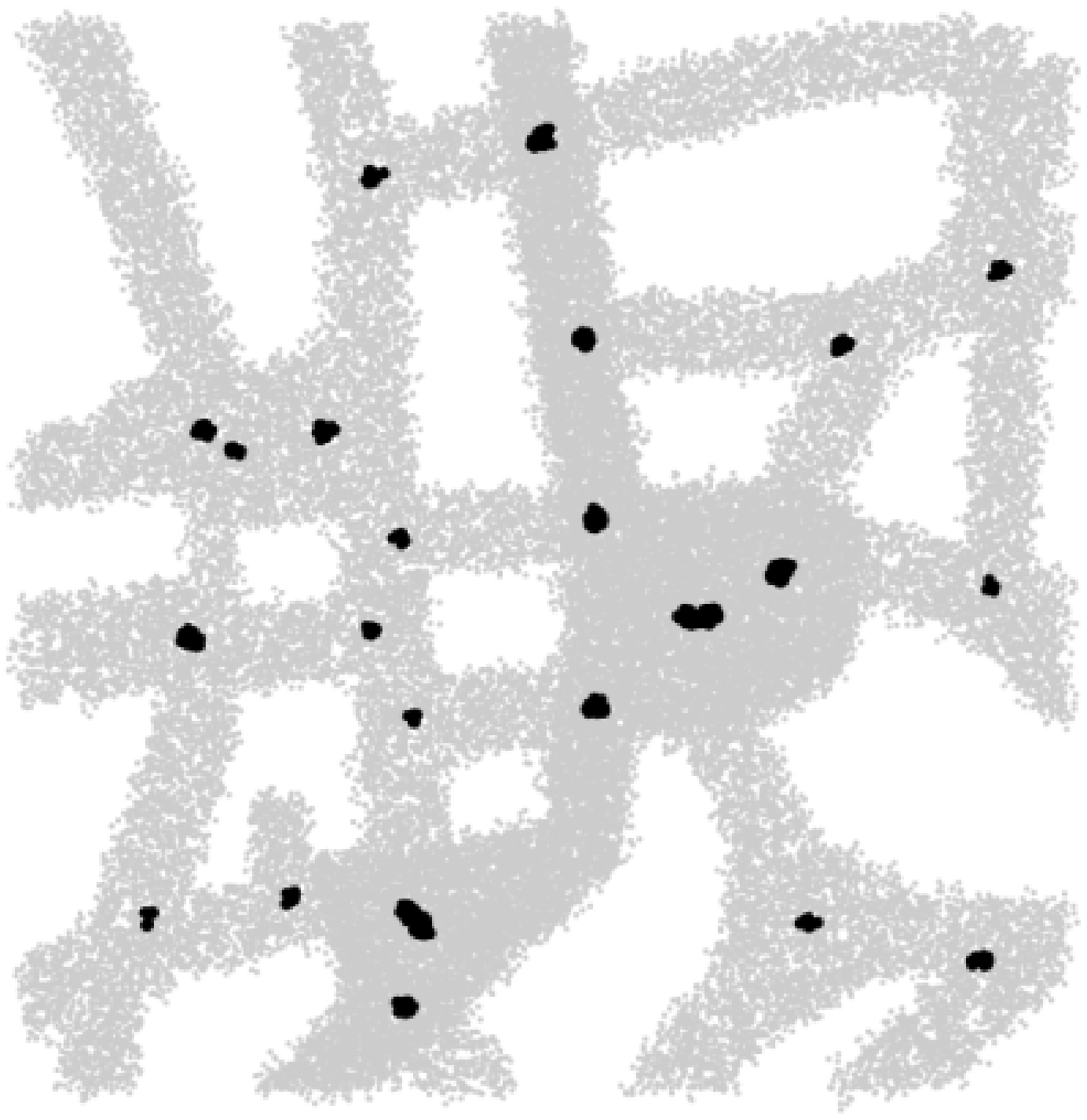}
  \caption{Identified 2-Voronoi and 3-Voronoi nodes.}
  \label{fig:cex:voronoi}
\end{figure}

\begin{figure}
  \centering
  \includegraphics[height=4cm]{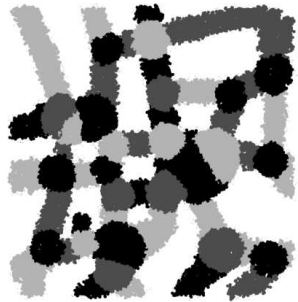}
  \caption{The final clustering.}
  \label{fig:cex:cluster}
\end{figure}

Figure~\ref{fig:cex:voronoi} shows the Voronoi sets $V_2$ and $V_3$.
One can clearly see the strips running between the boundaries and the
intersection cluster cores that are in the middle of intersections.
Finally, Figure~\ref{fig:cex:cluster} shows the clustering that is
computed by our method.  It consists of the intersection clusters
around the 3-Voronois, and street clusters in the remaining parts. The
geometric shape of the network area is reflected very closely, even
though the network had no access to geometric information.


\section{Conclusions}

In this paper we have described an integrated framework for the
deterministic self-organization of a large swarm of sensor nodes.
Our approach makes very few assumptions and is guaranteed to
produce correct results; the price is dealing with relatively
complex combinatorial structures such as flowers.
Obviously, stronger assumptions on the network properties,
the boundary structure or the distribution of nodes allow
faster and simpler boundary recognition; see our papers
\cite{cccg} and \cite{fkp-nbtrsn-04} for probabilistic ideas.

Our framework can be seen as a first step towards robust routing,
tracking and guiding algorithms. We are currently working on
extending our framework in this direction.


\bibliographystyle{abbrv}
\bibliography{references}



\end{document}